\documentstyle[a4,amssymb,amsthm,latexsym,12pt]{article}
\newcommand{\Med}[1]{\left\langle #1 \right\rangle}
\newcommand{\med}[1]{\langle #1 \rangle}

\newtheorem{teo}{Theorem}

\title{
Central limit theorem for fluctuations in the high temperature 
region of the Sherrington-Kirkpatrick spin glass model}

\author{
Francesco Guerra\footnote{\
e-mail: {\tt francesco.guerra@roma1.infn.it}} \\
{\small {\itshape Dipartimento di Fisica, Universit\`a di Roma `La Sapienza'}}
\\
{\small {\itshape and INFN, Sezione di Roma, Piazzale A. Moro 2, 00185 Roma, 
Italy}}\\
Fabio Lucio Toninelli\footnote{\ 
e-mail: {\tt f.toninelli@sns.it}} \\
{\small {\itshape Scuola Normale Superiore, Piazza dei Cavalieri 7, 56126 Pisa,
Italy}}\\
{\small {\itshape and Istituto Nazionale di Fisica Nucleare, Sezione di 
Pisa}}
} 

\date{\today}

\begin{document}

\maketitle


\begin{abstract}
In a region above the Almeida-Thouless line, where we are able to control 
the thermodynamic limit of the Sherrington-Kirkpatrick model and to prove
replica symmetry, we show that the fluctuations of the overlaps and of
the free energy are Gaussian, on the scale $1/\sqrt N$, for $N$ large.
The method we employ is based on the idea, we recently developed, of 
introducing quadratic coupling between two replicas. The proof makes use of the
cavity equations and of concentration
of measure inequalities for the free energy.
\end{abstract}

\clearpage

\section{Introduction}

We consider the mean field spin glass model introduced by Sherrington
and Kirkpatrick in \cite{sk}, \cite{sk2}, in the regime of high temperature or,
equivalently, of large magnetic field. On physical grounds, it is 
known that in this region the replica symmetric solution holds, as 
shown for example in \cite{MPV}, and references quoted there.
However, due to the very large fluctuations present in the model, 
it is difficult to give a mathematically rigorous description of 
this region. Rigorous works on this subject include \cite{alr},
\cite{guerra3}, \cite{comets}, \cite{fro1}, \cite{talaHT2}.
For other rigorous results concerning the model, we refer to
\cite{talagcorso}, \cite{talaRSB}, \cite{pastur}, \cite{guerra2}, \cite{gg}.

The method developed in \cite{talaHT2} by Michel Talagrand is 
particularly interesting. The starting point is the very deep 
physical idea that the phenomenon of replica symmetry breaking can 
be understood by studying the properties of the model under the application
of auxiliary interactions, which explicitly break replica symmetry.
In \cite{talaHT2}, this idea is employed to prove that the 
replica symmetric solution holds in a region, which ({\sl probably}) 
coincides with that found in the theoretical physics literature 
\cite{MPV}, i.e., up to the Almeida-Thouless critical line.

Recently \cite{noi} we proposed a different strategy, 
which consists in coupling two replicas of the system by means
of a term proportional to the square of the deviation of the
overlap from its replica symmetric value. 
In this way, we proved that 
replica symmetry holds in a region above the Almeida-Thouless line. 
In the same region, we obtained a control of the two-replica system, 
provided that the coupling parameter is small enough, and we showed that
the fluctuations of the overlap are  at most of order $1/\sqrt{N}$. In 
the present paper we prove that, in the same region of parameters,
the fluctuations of overlaps and free energy, when suitably rescaled,
have a Gaussian distribution when $N\to\infty$.
The main ingredients of the proof are the control of the 
thermodynamic limit obtained in \cite{noi}
and concentration of measure 
techniques inspired by Talagrand's works.
Then, by means of the cavity method, one can write self-consistent linear 
equations for the characteristic functions of the fluctuation variables,
which can be easily solved.

Previous results concerning limit theorems for fluctuations in the 
high temperature region of mean field spin glass models include
\cite{alr}, \cite{comets}, \cite{bovier}, \cite{talagcorso}, \cite{talalibro}.

This work is organized as follows: In Section \ref{modello},
we recall the main 
definitions of the model and introduce the overlap distribution structure.
In Section \ref{thtr}, we state the main results.
Two useful tools, i.e., exponential inequalities and 
the cavity method, are briefly outlined in Sections \ref{expon} and 
\ref{sez:cavita}.
In Sections \ref{centrale} and \ref{sec:fluttF}, we prove the central limit 
theorem for overlap and free energy fluctuations, respectively. Finally, 
Section \ref{conclusioni} is dedicated to a short outlook about open 
problems and further developments.

\section{The model}

\label{modello}

The generic configuration of the Sherrington-Kirkpatrick (SK) model is 
determined 
by the $N$ Ising variables $\sigma_i=\pm1$, $i=1,2,\ldots,N$,
and the Hamiltonian is
\begin{equation}
\label{SK}
H_N(\sigma,h;J)=-\frac1{\sqrt N} \sum_{(i,j)} J_{ij} \sigma_i\sigma_j-
h\sum_i \sigma_i,
\end{equation}
where the sum $\sum_{(i,j)}$ runs over all the $N(N-1)/2$ distinct couples of 
sites.
The $J_{ij}$'s (quenched noise) are independent centered unit Gaussian 
variables ${\cal N}(0,1)$. 
The first term in (\ref{SK}) is a long range random two body interaction,
while the second represents the interaction with a fixed 
external magnetic field $h$. 
For a given temperature $1/\beta$ we can introduce the disorder dependent
partition function
\begin{equation}
\label{Z}
Z_N(\beta,h;J)=\sum_{\{\sigma\}} \exp(-\beta H_N(\sigma,h;J))
\end{equation}
and the auxiliary function
\begin{equation}
\label{alfa}
\alpha_N(\beta,h)=\frac1NE\ln Z_N(\beta,h;J),
\end{equation}
where $E$ denotes the average with respect to the external noise 
$J$.
Note that $\alpha_N(\beta,h)$ is the
quenched average of the free energy per spin, apart from the multiplicative
factor $-1/\beta$.

For later convenience, it is useful to generalize the model (\ref{SK})
by introducing a quenched random external magnetic field, which at every
site is an independent Gaussian variable of strength $x>0$.
In other words, the Boltzmann factor of the system becomes
\begin{eqnarray}
\label{BF}
\exp\left(\sqrt{\frac tN}\sum_{(i,j)} J_{ij} \sigma_i\sigma_j+
\sum_i (\beta h+\sqrt x J_i)\sigma_i\right),
\end{eqnarray}
where $J_i$ are i.i.d. ${\mathcal N}(0,1)$ random variables, independent 
of the $J_{ij}$'s.
We let $t=\beta^2$ in the two-body term. In the following, we 
always regard the system as depending on the parameters $t,x, \beta h$.
In analogy with Eqs. (\ref{Z}), (\ref{alfa}), we define the disorder 
dependent partition function $Z_N(t,x,h;J)$ and the auxiliary function
$$\alpha_N(t,x,h)=\frac1NE\ln Z_N(t,x,h;J).$$
Here, $E$ denotes averaging with respect to $J_{ij}$ and $J_i$.
For simplicity of notations, here and in the following we write the 
argument $h$ instead of $\beta h$.

Let us consider a countably infinite number of independent copies (replicas) of
the system, whose spin variables $\sigma^a_i$ are distributed,
for fixed $J$, according to the product state 
$$\Omega_J=\Omega^{1}_J\Omega^{2}_J\ldots,$$
where $\Omega_J\equiv\Omega_J^{N,t,x,h}$ denotes the Gibbs state 
associated to the Boltzmann factor (\ref{BF}).
Each replica is subject to the same quenched noise.
The ``real replica'' approach has already been exploited in a number of papers 
\cite{guerra1}, \cite{guerra4}, \cite{ac}, \cite{parisi}.

The overlap between two replicas $a,b$ is defined as
$$q_{ab}(\sigma^{a},\sigma^{b})=\frac1N\sum_i\sigma^{a}_i\sigma^{b}_i,$$
with the obvious bounds
$$-1\leq q_{ab}\leq 1.$$
For a generic smooth function $F$ of the overlaps, we define the $\langle
.\rangle$ average as
$$\med{F(q_{12},q_{13},\ldots)}=E \Omega_J(F(q_{12},q_{13},\ldots)).$$
Note that the average over disorder introduces correlations between different
groups of replicas, which would be independent under the Boltzmann average
$\Omega_J$. For example,
$$\Omega_J(q_{12}q_{34})=\Omega_J(q_{12})\Omega_J(q_{34})$$
but
$$\med{q_{12}q_{34}}\neq\med{q_{12}}\med{q_{34}}.$$

\section{The high temperature region and the main results}

\label{thtr}

In this Section, we recall the results of \cite{noi} and
state limit theorems for fluctuations, in
the region where we prove that replica symmetry holds, i.e.,
$$\lim_{N\to\infty}\alpha_N(t,x,h)=\bar\alpha(t,x,h).$$
$\bar\alpha(t,x,h)$ is the replica-symmetric free energy \cite{sk}, \cite{sk2}
$$\bar\alpha(t,x,h)=\ln2+\int 
\ln\cosh( \beta h+z\sqrt{t\, \bar q+x})\,d\mu(z)+\frac t4(1-\bar q)^2,$$
$\bar q$ is the Sherrington-Kirkpatrick order parameter, defined as 
the unique \cite{guerra1} solution of 
$$\bar q=\bar q(t,x,h)=\int \tanh^2( \beta h+z\sqrt{t\,\bar q+x})\,d\mu(z)$$
and $d\mu(z)$ is the centered unit Gaussian measure.

In \cite{noi} we proved the following: 
Consider the auxiliary function $\tilde\alpha_N$, dependent on the 
parameter $\lambda\geq0$
$$\tilde \alpha_N(t,x,h;\lambda)=\alpha_N(t,x,h)+
\frac1{2N}E\,\ln \Omega_{t,x,h}\left(e^{N\,\frac\lambda2(q_{12}-\bar q)^2}
\right)$$
and the trajectory in the $(t,x)$ plane 
\begin{equation}
\label{ASDF}
\Gamma=(t',x_{t'})\equiv(t',x+\bar q (t-t'))\equiv (t',x_0-\bar q\,t'),
\hspace{1cm}0\leq t'\leq t
\end{equation}
where $x_0=x+\bar q\, t$ and $\bar q=\bar q(t,x,h)=\bar q(t',x_{t'},h)$.
Notice that $\tilde \alpha_N$ equals $\alpha_N$ for $\lambda=0$.
Given $x_0,h$ there exists a value $t_c(x_0,h)$, such that
\begin{equation}
\label{ris1}
\left|\bar\alpha(t',x_{t'},h)-\tilde\alpha_N(t',x_{t'},h;\lambda)\right|
\leq \frac kN
\end{equation}
for some constant $k$, uniformly in the triangular region
\begin{equation}
\label{triang}
0\leq t'+\lambda\leq \bar t<t_c(x_0,h).
\end{equation}
In the same region, the overlaps self-average around the 
value $\bar q$:
\begin{equation}
\label{ris2}
\med{(q_{ab}-\bar q)^2}\leq \frac kN.
\end{equation}
The critical value $t_c(x_0,h)$ is determined in the following way \cite{noi}:
Let
$$\Delta(x_0,h,\lambda_0)\equiv
\frac12\max_{\rho\in{\mathbb R}}\left(\int\ln(\cosh\rho+\tanh^2(\beta h + z 
\sqrt{x_0})\sinh\rho)
d\mu(z) -\rho\,\bar q-\frac{\rho^2}{2\lambda_0}\right),
$$
where $\lambda_0\ge0$.
Then, we define $t_c(x_0,h)$ such that, for any $\lambda_0\le t_c(x_0,h)$,
one has $$\Delta(x_0,h,\lambda_0)=0.$$
In the case of vanishing external field $x$$=$$h$$=$$0$, then
also $x_0$$=$$\bar q$$=$$0$ and $t_c=1$, the correct critical value.
As discussed in \cite{noi}, the region defined by (\ref{triang})
falls short of the 
Almeida-Thouless line, which is  the expected critical line.

In this paper, we investigate more precisely the behavior of fluctuations 
of physical quantities around the replica symmetric value.
First of all, we give a central limit-type theorem for the rescaled overlaps
$$\xi_{ab}^N=\sqrt N (q_{ab}-\bar q),$$
showing that they behave as centered Gaussian variables
characterized by a non-diagonal correlation matrix. 
Notice that, thanks to (\ref{ris2}), one has the following bound for the 
second moment of the rescaled overlap fluctuations:
\begin{equation}
\label{(*)}
\Med{(\xi_{ab}^N)^2}\le k.
\end{equation}
\begin{teo}
\label{teoflutQ}
If $t<t_c(x_0,h)$, the rescaled overlaps $\xi^N_{ab}$ tend in distribution, 
for $N\to\infty$,
to jointly Gaussian variables $\xi_{ab}$, with covariances
\begin{eqnarray}
\nonumber
&&\med{\xi^2_{ab}}=A(t,x,h)\\\nonumber
&&\med{\xi_{ab}\xi_{ac}}=B(t,x,h)\\\nonumber
&&\med{\xi_{ab}\xi_{cd}}=C(t,x,h),
\end{eqnarray}
where $b\neq c$, $c\neq a,b$ and $d\neq a,b$.
$A,B$ and $C$ are explicitly given by
\begin{eqnarray}
\label{covarianze1}
&&A(t,x,h)=(1+2R+4R^2)Y+c_0 R^2\\
&&B(t,x,h)=(1+4R)RY+c_0R^2\\\label{covarianze3}
&&C(t,x,h)=4R^2Y+c_0R^2,
\end{eqnarray}
where
\begin{eqnarray}
\nonumber
&&Y(t,x,h)=\frac1{Y_0^{-1}-t}\\\nonumber
&&R(t,x,h)=\frac{d_0}{Y_0^{-1}+2d_0-t}
\end{eqnarray}
and $Y_0(x_0,h),c_0(x_0,h)$ and $d_0(x_0,h)$ are chosen in such a way that
$A,B,C$ satisfy the initial conditions
\begin{eqnarray}
\nonumber
&&A(0,x_0,h)=1-\bar q^2\\\nonumber
&&B(0,x_0,h)=\bar q-\bar q^2\\\nonumber
&&C(0,x_0,h)=\int\tanh^4(z\sqrt {x_0}+\beta h)\,d\mu(z)-\bar q^2.
\end{eqnarray}
In particular, one has
\begin{eqnarray}
\label{y0}
Y_0=\int \cosh^{-4}(z\sqrt{t\,\bar q+x}+\beta h)\,d\mu(z)
\end{eqnarray}
\end{teo}
Recently, an analogous 
result was proved independently by Talagrand \cite{talalibro}, who
computed the $N\to\infty$ limit for all moments of the $\xi$ variables. 

The expressions for $A,B$ and $C$ were first given by Guerra in \cite{guerra1}.
For $h=x=0$, the limit Gaussian variables are not correlated and have
variance $1/(1-t)$, which is a well known result \cite{alr}, \cite{comets}.

Let us consider now free energy fluctuations. Aizenman, Lebowitz and Ruelle 
\cite{alr} proved that in the case of zero external 
field and $t<1$, the variable 
$$\ln Z_N-\ln EZ_N$$
tends to a shifted Gaussian random variable whose variance
diverges at  $t=1$. In the general case the situation is quite different
and the following theorem holds:
\begin{teo}
\label{fluttZ}
Let
$$\hat f_N(t,x,h;J)\equiv\sqrt N\left(\frac{\ln Z_N(t,x,h;J)}N-
\bar\alpha(t,x,h)\right).$$
If $t<t_c(x_0,h)$ then
$$\hat f_N(t,x,h;J)\stackrel{d}{\longrightarrow} 
{\mathcal N}(0,\sigma^2(t,x,h)),$$
where 
$$\sigma^2(t,x,h)={\mbox Var}\left(\ln\cosh(z\sqrt{t\,\bar q+x}+\beta h)\right)-
\frac{\bar q ^2 t}2$$
Here, $Var(.)$ denotes the variance of a random variable and $z={\mathcal N}
(0,1)$.
\end{teo}

Notice that fluctuations of the extensive free energy 
$\ln Z_N$ are of order $1$ at zero external field and of order
$\sqrt N$ otherwise.

\section{Exponential suppression of overlap fluctuations}

\label{expon}

General arguments based on concentration of measure 
\cite{talagcorso}, \cite{newlook}, \cite{talaconc}
show that the fluctuations of the free energy $1/N\ln Z_N$ around its 
mean value $\alpha_N$ are exponentially suppressed as $N$ grows. 
Indeed, one has the following \cite{talagcorso}
\begin{teo}
\label{teoconcentr}
For any $u>0$, 
\begin{equation}
\label{eqconcentr}
{P}\left(\left|\frac1N\log Z_N(t,x,h;J)-\alpha_N(t,x,h)\right|
\geq u\right)\leq \exp(-N K u^2),
\end{equation}
where
$$K=\frac 1{t+2x}.$$
\end{teo}
This, in connection with the results of \cite{noi}, allows to obtain a
strong control on the fluctuations of the overlaps
(we learned this nice argument in \cite{talaHT2}): First of all, the same 
argument leading to Theorem \ref{teoconcentr} shows that
$$P\left(\left|\frac1{2N}\ln\Omega_{t,x,h}
\left(e^{N\frac\lambda2(q_{12}-
\bar q)^2}\right)-\tilde\alpha_N(t,x,h,\lambda)+\alpha_N(t,x,h)\right|\geq 2u
\right)\leq \exp(-N K u^2).$$
Therefore, thanks to Eq. (\ref{ris1}), with probability at least 
$1-\exp(-N K u^2)$ one has
$$\Omega_{t,x,h}\left(e^{\frac\lambda2N(q_{12}-\bar q)^2}\right)
\leq e^{4Nu+2 C}$$
for $\lambda\leq\bar \lambda<t_c(x_0,h)-t$. Then, by Tchebyshev's inequality
$$\Omega_{t,x,h}\left(\chi_{\{|q_{12}-\bar q|\geq v\}}\right)\leq
e^{-\frac{\bar \lambda}2Nv^2}\Omega_{t,x,h}\left(e^{\frac{\bar \lambda}2N
(q_{12}-\bar q)^2}\right)\leq e^{N(4u-\frac{\bar \lambda}2v^2)+2C}$$
and, choosing $u=\bar \lambda v^2/16$, one has
$$\Omega_{t,x,h}\left(\chi_{\{|q_{12}-\bar q|\geq v\}}\right)\leq
e^{-N\frac{v^2\bar \lambda}4+2C},$$
The estimate we are looking for easily follows:
\begin{eqnarray}
\label{???}
E\,\Omega_{t,x,h}\left(\chi_{\{|q_{12}-\bar q|\geq v\}}\right)\leq
e^{-N\frac{v^2\bar \lambda}4+2C}+e^{-N\, K\frac{\bar\lambda^2 v^4}{256}}.
\end{eqnarray}
Of course, this is much more than just self-averaging of the overlaps.

\section{The cavity method}

\label{sez:cavita}

The cavity method allows to
express thermal averages of quantities defined on the $N$-spin system
as functions of averages on the system with $N-1$ spins, at a 
slightly different temperature. This method has been widely applied 
both in the theoretical physics literature \cite{MPV} and in the mathematical 
physics one (see, for 
instance, \cite{guerra3}, \cite{talalibro}, \cite{talaRSB}, \cite{guerra4}).

Introduce the following definitions:
\begin{eqnarray}
\nonumber &&t'=t\,(1-N^{-1})\\\nonumber
&&\sigma^a=(\eta^a,\epsilon^a),\hspace{1cm}\eta^a
\in \{-1,1\}^{N-1},\hspace{1cm}\epsilon^a=\sigma^a_N =\pm1\\\nonumber
&&J=J_{N}\\\nonumber
&&g_i= J_{N\,i},\hspace{1cm}i=1,\ldots,N-1\\\nonumber
&&\Omega'(.)=\Omega_{N-1}^{t'}(.)
\end{eqnarray}
The cavity equations consist in the identity
\begin{equation}
\label{cavita2}
\Omega^{t,x,h}_{N}\left(f(\sigma^1,\ldots,\sigma^k)\right)=
\frac{\Omega'\left(Av\,
f(\eta^1,\epsilon^1,\ldots,\eta^k,\epsilon^k){\Psi^{(k)}}
\right)}{\Omega'\left(Av\,{\Psi^{(k)}}\right)},
\end{equation}
where $Av$ denotes the average over the spin variables 
$\epsilon^a$ and
\begin{equation}
\label{cavita3}
{\Psi^{(k)}}\equiv
\exp \sum_{a=1}^k\epsilon^a\left(\sqrt{ t/N} g\,\eta^a+\sqrt x J +\beta
h\right).
\end{equation}
$g\,\eta^a$ denotes the scalar product $\sum_{i=1}^{N-1}g_i \eta^a_i$.

\section{Limit theorem for overlap fluctuations}

\label{centrale}

\label{sec:flucQ}

To prove Theorem \ref{teoflutQ},
it suffices \cite{shiri} 
to show that for any integer $s$, the characteristic function 
$$\phi_N^{t}(u)=\Med{\exp i\,u\,\xi^N}
\equiv \Med{\exp i\,\sum_{(a,b)}u_{ab}\,\xi^N_{ab} },\hspace{1cm}1\le a<b\le s$$
converges for $N\to\infty$ to
\begin{equation}
\label{post}
\phi^t(u)=\exp\left\{-\frac12\, (\hat L u,u)\right\},
\end{equation}
where $(.\,,.)$ denotes scalar product and $\hat L$ is the $s(s-1)/2\times
s(s-1)/2$ dimensional matrix of elements
\begin{eqnarray}\nonumber
&&L_{(ab),(ab)}=A(t,x,h)\\\nonumber
&&L_{(ab),(ac)}=B(t,x,h)\\\nonumber
&&L_{(ab),(cd)}=C(t,x,h).
\end{eqnarray}
The idea of the proof is 
to obtain a set of closed linear differential equations for $\phi_N^{t}(u)$,
which determine uniquely the solution as (\ref{post}), for $N\to\infty$.
Some of the calculations involved in the proof are quite long, although 
straightforward, and are therefore just sketched.

First of all, we explain how the cavity equations (\ref{cavita2}), 
(\ref{cavita3}) can be simplified in the region where (\ref{???}) holds.
Following \cite{talaRSB}, we introduce some notations, letting
$\Omega(.)\equiv\Omega_N^{t,x,h}(.)$ and $\Omega'(.)
\equiv\Omega_{N-1}^{t',x,h}(.)$.
Moreover, we define
\begin{eqnarray}\nonumber
&&b=\Omega'(\eta)\in {\mathbb R}^{N-1}\\\nonumber
&&\dot\eta^a=\eta^a-b\\\nonumber
&&X=\sqrt{t/N}g\,b+\sqrt x J+\beta h\\\nonumber
&&\Psi^{(k)}_0=\exp (X\sum_{a=1}^k \epsilon^a)\\\nonumber
&&f(\sigma^1,\ldots,\sigma^k)=f(\eta^1,\epsilon^1,\ldots,\eta^k,\epsilon^k).
\end{eqnarray}
Then, the following holds \cite{talaRSB}:
\begin{teo}
\label{cavita'}
\begin{eqnarray}
\label{papjiro}
E\Omega\left( f(\sigma^1,\ldots,\sigma^k)\right)&=&
E\frac 1{\cosh^k X}\Omega'\left(Av\, f\, \Psi^{(k)}_0\right)
\\\label{pap2}
&&+t\,E\frac 1{\cosh^k X}\Omega'\left(Av f\,\Psi^{(k)}_0
\sum_{1\leq a<c\leq k}\epsilon^a\epsilon^c\frac{\dot\eta^a\dot\eta^c}N
\right)\\\label{pap4}
&&+t\,E\frac 1{\cosh^k X}\Omega'\left(Av f\,\Psi^{(k)}_0
\sum_{1\leq a\neq c\leq k}\epsilon^a\epsilon^c\frac{\dot\eta^c b}N\right)\\\label{pap5}
&&-k\,t\,E\frac{\tanh X}{\cosh^k X}\Omega'\left(Av f\,\Psi^{(k)}_0
\sum_{a=1}^k\epsilon^a\frac{\dot\eta^a b}N\right)+S
\end{eqnarray}
and the ``error term'' $S$ can be estimated as
\begin{eqnarray}\nonumber
|S|\leq w_k(t,x,h)E\Omega'\left(Av |f|\left(\sum_{a=1}^{k+1}
\left(\frac{\dot\eta^a b}N\right)^2+\sum_{1\leq a<c\leq k+2}
\left(\frac{\dot\eta^a \dot \eta^c}N\right)^2\right)\right),
\end{eqnarray}
where $w$ is a smooth function of its arguments, independent of $N$.
\end{teo}
Note that, with respect to Theorem 3.2 in \cite{talaRSB}, 
the last sum in the r.h.s. is performed on $a<c$ instead of $a\leq c$.
However, the proof of Theorem \ref{cavita'} proceeds
exactly as in \cite{talaRSB}.

Theorem \ref{cavita'}  is 
a sort of Taylor expansion of the cavity equations
around $\eta^a=b$. This turns out to be particularly useful in the 
region where Eq. (\ref{???}) holds, since in this case
$\dot\eta_a$ is small with large probability, and $S$ vanishes for 
$N\to\infty$, as we explain in the following.

In order to prove Theorem \ref{teoflutQ}, we first exploit symmetry between 
sites to write
\begin{eqnarray}
\label{aaa}
&&\partial_{u_{rr'}}\phi_N^t(u)=i\Med{\xi^N_{rr'}\,e^{i\,u\,\xi^N}}
=i\sqrt N\Med{(\sigma^r_N\sigma^{r'}_N-\bar q)\,e^{i\,u\,\xi^N}}\\
\label{bbb}
&&\varphi_N^{(a)\,t}(u)\equiv i\Med{\xi^N_{a,s+1}\,e^{i\,u\,\xi^N}}=
i\sqrt N\Med{(\sigma^a_N\sigma^{s+1}_N-\bar q)\,e^{i\,u\,\xi^N}}\\
\label{ccc}
&&\psi_N^t(u)\equiv i\Med{\xi^N_{s+1,s+2}\,e^{i\,u\,\xi^N}}
=i\sqrt N\Med{(\sigma^{s+1}_N\sigma^{s+2}_N-\bar q)\,e^{i\,u\,\xi^N}},
\end{eqnarray}
and then employ the cavity equations 
to express these quantities as functions of $\phi,\varphi,\psi$ themselves.
For instance, apply Theorem \ref{cavita'} to the r.h.s. of Eq. (\ref{aaa})
and consider the term arising from 
(\ref{papjiro}). After averaging on the dichotomic variables $\epsilon$, one is left with
\begin{eqnarray}
\label{left}
&&i\,(\sqrt N-i\,\sum_{(a,b)}u_{ab}\,\bar q)E\left\{(\tanh^2 X-\bar q)
\Omega'\exp\left(i\,u'\,\xi^{N-1}\right)\right\}+
\\\nonumber 
&&-\,u_{rr'}E\left\{(1-\bar q\tanh^2 X)
\Omega'\exp\left(i\,u'\,\xi^{N-1}\right)\right\}
\\\nonumber
&&-(1-\bar q)\sum_{a\neq r,r'}(u_{ar}+u_{ar'})
E\left\{ \tanh^2 X\,\Omega'\exp\left(i\,u'\,\xi^{N-1}\right)\right\}
\\\nonumber
&&-\,\sum_{(c,d)\,c,d\neq r,r'} u_{cd}\,
E\left\{ \tanh^2 X\left(\tanh^2 X-\bar q\right)
\Omega'\exp\left(i\,u'\,\xi^{N-1}\right)\right\}+o(1),
\end{eqnarray}
where $u'=u\sqrt{1-1/N}$. 
The term $o(1)$ arises when  $\exp\,(iu/\sqrt N)$ is expanded
around $u$$=$$0$ and the terms of order $u^2$ or higher are neglected.
Indeed, one has
\begin{eqnarray}
\nonumber
&&\left|\sqrt{N}\Med{(\sigma^r_N\sigma^{r'}_N-\bar q)
\, e^{i u'\xi^{N-1}}\left(e^{i\sum_{(a,b)}\frac{u_{ab}}{\sqrt N}
(\sigma^a_N\sigma^b_N-\bar q)}-1-i\sum_{(a,b)}\frac{u_{ab}}{\sqrt N}
(\sigma^a_N\sigma^b_N-\bar q)\right)}\right|\\\nonumber
&&\le2\sqrt N \left|e^{i\sum_{(a,b)}\frac{u_{ab}}{\sqrt N}
(\sigma^a_N\sigma^b_N-\bar q)}-1-i\sum_{(a,b)}\frac{u_{ab}}{\sqrt N}
(\sigma^a_N\sigma^b_N-\bar q)\right|={O}(N^{-1/2}).
\end{eqnarray}
Now, rewrite $E$ as ${E\, E}_g$, where ${E}_g$ denotes the average only with 
respect to the random variables $J$ and $g_i, i=1,\ldots,N-1$, and notice that 
$J, g_i$ do not appear in the thermal average  $\Omega'$.
Computation of $E_g(\ldots)$
would be simpler if, instead of $X$, there were
$$\bar X\equiv \sqrt{t/N}g\,\bar b+\sqrt x J+\beta h,$$
where 
$$\bar b\equiv \frac b{||b||}\sqrt{N\bar  q}.$$
Of course, one has
$$
\bar X\stackrel{d}{=}z\sqrt{t\,\bar q+x}+\beta h,
$$
where $z$ is a standard unit Gaussian variable and equality holds in 
distribution so that, for instance, 
$$E_g\tanh^2 \bar X=\bar q.$$
The idea is, therefore, to expand around $X=\bar X$.
As a preliminary fact, notice that the second moment of the random variable
$(b-\bar b)$ is bounded uniformly in $N$. Indeed, 
\begin{eqnarray}
&&E||b-\bar b||^2=E(||b||-\sqrt{N\bar q})^2\le\frac1{\bar q N}E(||b||^2-
N\bar q)^2\\
\label{O1}
&&=\frac1{\bar q}E\Omega'(\xi^{N-1}_{12}\xi^{N-1}_{34})+O(1/N)=O(1),
\end{eqnarray}
thanks to Eq. (\ref{(*)}).
As an example, let us examine in detail the first term in 
(\ref{left}), that is,
\begin{eqnarray}\label{24}
&&i\sqrt N E\,E_g (\tanh^2 X) \,\Omega'
\exp\left(i\,u'\xi^{N-1}\right)-i \,\bar q\sqrt N E\,\Omega'
\exp\left(i\,u'\xi^{N-1}\right).
\end{eqnarray}
By a simple second order Taylor expansion and an 
integration by parts on the Gaussian noise $g$, one finds
\begin{eqnarray}
\label{primo}
&&E_g\, \tanh^2 X=E_g\,\tanh^2 \bar X+\frac tN (b-\bar b)\bar b\,E_g\,
\left.\partial^2_x \tanh^2x\right|_{x=\bar X} \\
\label{secondo}
&&+\frac{t}{2N}E_g\, \partial^2_x \left.\tanh^2x\right|_{x=\bar X+
\theta(X-\bar X)}\left(g(b-\bar b)\right)^2\\
\label{terzo}
&&=\bar q+\frac t{2N}(b-\bar b)(b+\bar b)E_g\,\partial^2_x \tanh^2\bar X
\\
\label{quarto}
&&+\frac t{2N}||b-\bar b||^2E_g\left(\partial^2_x\left.\tanh^2x\right|
_{x=\bar X+\theta(X-\bar X)}-\partial^2_x\tanh^2\bar X\right)\\
\label{quinto}
&&+\frac{t^2}{2N^2}E_g\,\left.\partial^4_x\tanh^2x
\right|_{x=\bar X+\theta(X-\bar X)}[(b-\bar b)(\bar b+\theta(b-\bar b))]^2
\end{eqnarray}
where $0\le\theta\le1$.
Let analyse each term separately. Recalling the definitions of $b$ and 
$\bar b$, the second term in (\ref{terzo}) equals
\begin{eqnarray}
&&\frac t{2N}\Omega'(\eta^{s+1}\eta^{s+2}-N\bar q)
\int d\mu(z)\partial^2_x\tanh^2(\beta h+z\sqrt{t\,\bar q+x})\\
&&=\frac t{\sqrt N}\Omega'(\xi^{N-1}_{s+1,s+2})(3Y_0+2\bar q-2)
+O(1/N),
\end{eqnarray}
where $Y_0$ was defined in (\ref{y0}). 
Another application of Taylor's expansion and integration
by parts, together with Cauchy-Schwarz inequality and the fact that 
the derivatives of the function $\tanh^2(x)$ are bounded, shows that 
the terms (\ref{quarto}) and (\ref{quinto}) can be bounded by
$$\frac kN||b-\bar b||^2.$$
Therefore, using the estimate (\ref{O1}), the expression (\ref{24}) reduces to
$$i\,t(3Y_0+2\bar q-2)E\,\Omega'\left[\xi_{s+1,s+2}\exp \left(i\,u'\,\xi^{N-1}
\right)\right]+O(N^{-1/2}),$$
and
$$i\sqrt NE\left\{(\tanh^2 X-\bar q)
\Omega' \exp\left(i\,u'\xi^{N-1}\right)\right\}=
\,t(2\bar q-2+3Y_0)\psi'+o(1),$$
where
$$\psi'\equiv\psi_{N-1}^{t'}(u').$$
The other terms in (\ref{left}) are much simpler than (\ref{24}), and 
can be dealt with in the same way.
Finally, the whole expression (\ref{left}) can be rewritten as
\begin{eqnarray}
\label{a}
&&t(2\bar q-2+3Y_0)\,\psi'-u_{rr'}(1-\bar q^2)\,\phi'\\\nonumber
&&-(\bar q-\bar q^2)\sum_{a\neq r,r'}(u_{ar}+u_{ar'})\phi'
-(Y_0-(1-\bar q)^2)\sum_{(c,d)\,c,d\neq r,r'} u_{cd}\,\phi'+o(1).
\\\nonumber
\end{eqnarray}
The steps leading to expression (\ref{a}) can be repeated with minor
changes for the remaining terms (\ref{pap2}) to (\ref{pap5}).
These terms, although they look more complicated than (\ref{papjiro}) at 
first sight, are actually simpler to treat, since a first (instead of second) 
order Taylor expansion around $X=\bar X$ is sufficient. This is due to the
presence of terms like $\dot \eta_a\dot\eta_b/N$ or  $\dot \eta_a b/N$,
which are with large probability small, thanks to (\ref{(*)}). Also in 
this case, one finds
that terms (\ref{pap2}) to (\ref{pap5}) give quantities linear 
in $\phi',\partial \phi',\varphi',\psi'$, apart from terms of order $o(1)$.
As for the ``error term'' $S$ which appears in Theorem \ref{cavita'}, 
one can easily check that it vanishes in the 
thermodynamic limit. This is a consequence of the exponential decay of overlap 
fluctuations, as expressed by (\ref{???}).

Next, we show that terms like $\phi_{N-1}^{t'}(u')$ or $\psi_{N-1}^{t'}(u')$
can be substituted by the same functions
calculated at $N,t,u$, apart from negligible error terms. Indeed, for instance,
\begin{eqnarray}\nonumber
\phi_N^t(u)&=&\Med{\exp \left(i\,u'\,\xi^{N-1}+
i\,u\,(\sigma^1_N\sigma^2_N-\bar q)/\sqrt N\right)}_t\\\nonumber
&=&\Med{\exp i\,u'\,\xi^{N-1}}_t(1+o(1))=\phi'+o(1).
\end{eqnarray}
In the last step, we used Theorem \ref{cavita'} in order to substitute
$t$ with $t'$. 
Therefore, Eq. (\ref{aaa}) reduces to a linear relation between
$\phi, \varphi$ and $\psi$, apart from a remainder which becomes irrelevant
in the thermodynamic limit. 
In the same way, one sees that also Eqs. (\ref{bbb}), (\ref{ccc}) yield 
linear equations for $\phi,\varphi,\psi$. 
Putting everything together, in the thermodynamic limit one has a 
set of coupled linear differential equations of the form
\begin{equation}
\label{elegante}
{\bf \Phi}^t(u)= \phi^t(u)\,{\bf v}(u)+t\,\hat M\,{\bf\Phi}^t(u)
\end{equation}
where ${\bf \Phi}^t(u)$ is the vector 
$${\bf \Phi}^t(u)=(\partial_{u_{12}}\phi^t(u),\ldots,\partial_{u_{s-1,s}}\phi^t(u),
\varphi^{(1)\,t}(u),\ldots,\varphi^{(s)\,t}(u),\psi^t(u)).$$
${\bf v}(u)$ is a vector whose components are homogeneous linear
functions of  the variables $u$, while $\hat M$ is a real square matrix with elements depending 
on $\bar q, Y_0$ alone.
We do not report here the explicit expressions of ${\bf v}(u)$ and $\hat M$, 
which are quite complicated. However, it is instructive to check that, for 
instance, the term (\ref{a}) is in agreement with this structure. In fact, 
the coefficient of $\phi'$ is a homogeneous linear function of the $u$ 
variables, while the coefficient of $\psi'$ is linear in t and depends only on
$Y_0$ and $\bar q$. As will be clear in the following, only the 
structure (\ref{elegante}), and not the specific form of ${\bf v}$ 
and $\hat M$, are needed to conclude the proof of the theorem.

Assume at first that the matrix $(1-t \,\hat M)$ is invertible, 
which in principle can fail only for a finite number of values of $t$, 
since $\hat M$ is finite dimensional.
In this case, Eq. (\ref{elegante}) can be reduced to a first order 
differential system in normal form:
\begin{equation}
\label{..}
{\bf \Phi}^t(u)=\phi^t(u)(1-t\,\hat M)^{-1}{\bf v}(u),
\end{equation}
which can be easily integrated.
The most general solution for $\phi^t(u)$, compatible with the initial
condition $$\phi^t(0)=1,$$ is of the form 
\begin{equation}\label{form}
\phi^t(u)=\exp\left\{-\frac12(\hat K u,u)+(p, u)\right\},
\end{equation}
where $p$ is some $s(s-1)/2$ dimensional $u$-independent vector,
and $\hat K$ is a $s(s-1)/2\times s(s-1)/2$ real symmetric positive definite 
matrix.
The symmetry and non negativity of $\hat K$ derive from the obvious property
of symmetry among replicas, and from the bound
$$\left|\phi^t(u)\right|\le1,$$
which holds for any characteristic function.
The quadratic dependence on $u$ of the 
exponent of $\phi^t(u)$ stems from the linear dependence of the components 
of ${\bf v}(u)$.
Clearly, Eq. (\ref{form}) means that the random variables $\{\xi^N_{ab}\}$
converge to some Gaussian process $\{\xi_{ab}\}$. Moreover, it turns out that 
the identification
$$p=0$$ and $$\hat K=\hat L$$ are straightforward. Indeed, it was 
shown by Guerra in \cite{guerra1} that, if the limit process is Gaussian,
then it is centered and its covariance function is exactly $\hat L$.

In order to conclude the proof, it remains to show convergence of the 
characteristic function for those possible values $\tilde t$ where 
$(1-t \,\hat M)$ is singular.
For any $\delta>0$ one can write
$$\phi_N^{\tilde t}(u)=\phi_N^{\tilde t-\delta}(u)+\delta\,\partial_t
\left.\phi_N^t\right|_{t=\tilde t-\theta_N\,\delta},$$
where $0<\theta_N<1$.
After a straightforward computation one finds that
$$\partial_t\phi_N^t=\frac12 \Med{e^{i\,u\xi^N}\,\left(
\sum_{(a,b)}(\xi^N_{ab})^2-s\sum_{a=1}^s (\xi^N_{a,s+1})^2+\frac
{s(s+1)}2(\xi^N_{s+1,s+2})^2\right)}.$$
By exploiting the uniform bound (\ref{(*)}) and the arbitrariness of $\delta$, 
one finds therefore that the theorem holds also for $t=\tilde t$.
$\Box$

\section{Fluctuations of the free energy}

\label{sec:fluttF}

In order to prove Theorem \ref{fluttZ},
we show that the characteristic function of $\hat f_N$ converges
to that of ${\mathcal N}(0,\sigma^2(t,x,h))$, i.e., 
$$\lim_{N\to\infty} E\,e^{i\,u\, \hat f_N(t,x,h)}=
e^{-\frac{u^2}2 \sigma^2(t,x,h)}.$$
Define
\begin{eqnarray}\nonumber
&&\bar \alpha(t')=\bar \alpha(t',x_{t'},h)\\\nonumber
&&\zeta_N(t')= \frac{\ln Z_N(t',x_{t'},h;J)}N,
\end{eqnarray}
where $x_{t'}$ is defined in Eq. (\ref{ASDF}).
The characteristic function of $\hat f_N$ can be written as
\begin{eqnarray}
\label{party0}
E\,e^{i\,u \hat f_N(t,x,h)}&=&E\,e^{i\,u \hat f_N(0,x_0,h)}
+i\,u\, E\int_0^t\,e^{i\,u \hat f_N(t',x_{t'},h)}
\,\frac{d}{dt'} \hat f_N(t',x_{t'},h)
dt'.
\end{eqnarray}
Since
\begin{eqnarray}\nonumber
&&\frac d{dt'}\bar\alpha(t')=\frac14(1-\bar q)^2\\\nonumber
&&\frac d{dt'}\zeta_N(t')=\frac1{2\sqrt{t'} N^{3/2}}\sum_{(i,j)}
J_{ij}\Omega_{t'}(\sigma_i\sigma_j)-\frac{\bar q}{2N\sqrt {x_{t'}}}\sum_i J_i
\Omega_{t'}(\sigma_i),
\end{eqnarray}
one finds through integration by parts that
\begin{eqnarray}
\label{party}
E \left\{e^{i\,u\,\zeta_N(t')}\frac d{dt'}
\zeta_N(t')\right\}&=&\frac14E\left\{e^{i\,u\,\zeta_N(t')}
[1-\Omega_{t'}(q_{12}^2)-2\bar q\,(1-\Omega_{t'}(q_{12}))]\right\}
\\\nonumber
&&+\frac {i\,u}{4N}E\left\{e^{i\,u\,\zeta_N(t')}
[\Omega_{t'}(q_{12}^2)-2\bar q\,\Omega_{t'}(q_{12})-N^{-1}]\right\}.
\end{eqnarray}
By using (\ref{party}) in Eq. (\ref{party0}), one finds 
\begin{eqnarray}
\label{*}
E\,e^{i\,u \hat f_N(t,x,h)}&=&E\,e^{i\,u \hat f_N(0,x_0,h)}+\frac{u^2\bar q^2}4
E\int_0^t\,e^{i\,u \hat f_N(t')}\,dt'
\\\nonumber
&&-\frac{u^2}{4N}\int_0^t E\,e^{i\,u \hat f_N(t')}
\left(\Omega_{t'}(\xi_{12}^2)-1\right)\,dt'\\\nonumber
&&-\frac {i\,u}{4\sqrt N}\int_0^tE\, e^{i\,u \hat f_N(t')}
\Omega_{t'}(\xi_{12}^2)\,dt'.
\end{eqnarray}
At $t=0$, all sites are decoupled and the central limit theorem for i.i.d. 
random variables implies that 
\begin{equation}
\label{keeping}
\hat f_N(0,x,h)\stackrel{d}\longrightarrow {\mathcal N}(0,\sigma^2(0,x,h)).
\end{equation}
The last two terms in Eq. (\ref{*}) clearly vanish for 
$N\to\infty$. For instance,
$$N^{-\frac12}\left|E\,e^{i\,u \hat f_N(t')}\Omega_{t'}(\xi_{12}^2)\right|
\leq N^{-\frac12}E \Omega_{t'}(\xi_{12}^2)=O(N^{-\frac12}),$$
since
$$E \Omega_{t'}(\xi_{12}^2)=O(1)$$
for $t<t_c$.
Therefore, Eq. (\ref{*}) yields the following linear integral equation for 
the characteristic function:
$$E\,e^{i\,u \hat f_N(t,x,h)}=E\,e^{i\,u \hat f_N(0,x_0,h)}+
\frac{u^2\bar q^2}4
E\int_0^t\,e^{i\,u \hat f_N(t',x_{t'},h)}\,dt'+o(1),$$
whose solution is, keeping into account the initial condition (\ref{keeping}),
$$
E\,e^{i\,u \hat f_N(t,x,h)}=e^{-\frac{u^2}2 \sigma^2(t,x,h)}+o(1).
$$
$\Box$

Before concluding this Section, we wish to note that from Eq. (\ref{*}) one can
also obtain in a very simple way a well known result for free energy 
fluctuations at zero external field and $t<1$ \cite{alr, comets}, i.e.,
\begin{equation}
\label{noto}
\eta_N^t\equiv
\ln Z_N(t)-N\left(\ln2+\frac t4\right)\stackrel{d}{\longrightarrow} 
\hat Y_t-\frac14\ln\frac1{1-t},
\end{equation}
where $Y_t$ is a centered Gaussian random variable of variance
$$\frac12\left(\ln\frac1{1-t}-t\right).$$
Indeed, setting $u=\sqrt N s$ and $x=h=0$ in Eq. (\ref{*}), one 
obtains the equation
\begin{equation}
\label{**}
E\,e^{i\,s \,\eta^t_N}=1-\frac{u^2}{4}\int_0^t E\,e^{i\,s\, \eta_N^{t'}}
\left(\Omega_{t'}(\xi_{12}^2)-1\right)\,dt'
-\frac {iu}{4 }\int_0^tE\, e^{i\,s\, \eta_N^{t'}}
\Omega_{t'}(\xi_{12}^2)\,dt'.
\end{equation}
Since Theorem 1 implies, for vanishing external field and $t<1$,
$$E\left(\Omega_t(\xi^2_{12})-\med{\xi^2_{12}}\right)^2=
\med{\xi^2_{12}\,\xi^2_{34}}-\med{\xi_{12}^2}^2=
o(1),$$
Eq. (\ref{**}) yields 
$$E\,e^{i\,s \,\eta^t_N}=1-\frac{u^2}{4}\int_0^t E\,e^{i\,s\, \eta_N^{t'}}
\left(\frac1{1-t'}-1\right)\,dt'
-\frac {iu}{4 }\int_0^tE\, e^{i\,s\, \eta_N^{t'}}
\frac1{1-t'}\,dt'+o(1),$$
from which the result (\ref{noto}) easily follows.

\section{Conclusions and outlook}

\label{conclusioni}

We have employed the cavity method to prove a central limit theorem 
for the fluctuations of overlaps and free energy, in a region
above the Almeida-Thouless line. The key
ingredient was provided by the control of the coupled two replica system.
The open question remains to understand whether and how our method can be 
extended to the entire physically expected high temperature region.

In the case of vanishing external field, our method can be employed 
to obtain very detailed informations on the system in proximity of the 
critical point. In particular, one can obtain lower and upper bounds on
the overlap fluctuations, at $\beta=1$. We plan to report soon 
on this \cite{noinuovo}.

\vspace{.5cm}
{\bf Acknowledgments}

F. L. T. would like to thank Massimiliano Gubinelli for useful comments and
remarks.

This work was supported in part by MIUR 
(Italian Minister of Instruction, University and Research), 
and by INFN (Italian National Institute for Nuclear Physics).


\addcontentsline{toc}{section}{References}





\begin{thebibliography}{199}

\bibitem{sk} D. Sherrington, S. Kirkpatrick, 
{\sl Solvable model of a spin-glass}, 
Phys. Rev. Lett. {\bf 35}, 1792 (1975).

\bibitem{sk2} S. Kirkpatrick, D. Sherrington,
{\em Infinite-ranged models of spin-glasses}, 
Phys. Rev. B {\bf 17}, 4384 (1978).

\bibitem{MPV} M. M\'ezard, G. Parisi and M. A. Virasoro, {\sl Spin glass theory
and beyond}, World Scientific, Singapore (1987).

\bibitem{alr} M. Aizenman, J. Lebowitz and D. Ruelle, 
{\em Some rigorous results on the Sherrington-Kirkpatrick spin glass model}, 
Commun. Math. Phys. {\bf 112}, 3 (1987).

\bibitem{guerra3} F. Guerra, {\sl Fluctuations and thermodynamic variables
in mean field spin glass models}, in: {\sl Stochastic processes, physics and 
geometry, II}, S. Albeverio {\sl et al.}, eds., Singapore (1995).

\bibitem{comets} F. Comets, J. Neveu, 
{\sl The Sherrington-Kirkpatrick
model of spin glasses and stochastic calculus: the high temperature case},
Commun. Math. Phys. {\bf 166}, 549 (1995).

\bibitem{fro1} J. Fr\"ohlich, B. Zegarlinski, 
{\sl Some comments on the
Sherrington-Kirkpatrick model of spin glasses}, 
Commun. Math. Phys. {\bf 112}, 553 (1987).

\bibitem{talaHT2} M. Talagrand, {\sl On the high temperature region of the 
Sherrington-Kirkpatrick model}, C. R. A. S., to appear;
M. Talagrand, {\sl On the high temperature phase of the 
Sherrington-Kirkpatrick model}, Ann. Probab., to appear.

\bibitem{talagcorso} M. Talagrand, 
{\sl Mean field models for spin glasses:
a first course}, course given in Saint Flour in summer 2000, to appear.

\bibitem{talaRSB} M. Talagrand, 
{\sl Replica symmetry breaking and exponential 
inequalities for the Sherrington Kirkpatrick model}, 
Ann. Probab. {\bf 28}, 1018 (2000).

\bibitem{pastur} L. Pastur, M. Shcherbina, 
{\sl The absence of self-averaging
of the order parameter in the Sherrington-Kirkpatrick model}, 
J. Stat. Phys. {\bf 62}, 1 (1991).

\bibitem{guerra2} F. Guerra, 
{\sl About the overlap distribution in mean field
spin glass models}, 
Int. Jou. Mod. Phys. B {\bf 10}, 1675 (1996).

\bibitem{gg} S. Ghirlanda, F. Guerra, 
{\sl General properties of overlap 
distributions in disordered spin systems. Towards Parisi ultrametricity},
J. Phys. A {\bf 31}, 9149 (1998).

\bibitem{noi} F. Guerra, F. L. Toninelli, {\sl Quadratic replica coupling for 
the Sherrington-Kirkpatrick mean field spin glass model}, J. Math. Phys., 
{\bf 43}, 3704 (2002).

\bibitem{bovier} A. Bovier, I. Kurkova, M. L\"owe, 
{\sl Fluctuations of the free energy in the REM and the $p$-spin SK models},
Ann. Probab., to appear.

\bibitem{talalibro} M. Talagrand, 
{\sl Spin glasses: a challenge for mathematicians. Mean field models and 
cavity method}, Springer-Verlag, to appear.

\bibitem{guerra1} F. Guerra, {\sl Sum rules for the free energy in the mean 
field spin glass model}, Fields Institute 
Communications {\bf 30}, 161 (2001).

\bibitem{guerra4} F. Guerra, {\sl The cavity method in the mean field 
spin glass model. Functional representations of thermodynamic variables},
in: {\sl Advances in dynamical systems and quantum physics}, S. Albeverio 
{\sl et al.}, eds., Singapore (1995).

\bibitem{ac} M. Aizenman, P. Contucci, 
{\sl On the stability of the quenched state in mean field spin glass models}, 
J. Stat. Phys. {\bf 92}, 765 (1998).

\bibitem{parisi} G. Parisi, {\sl On the probabilistic formulation
of the replica approach to spin glasses}, cond-mat/9801081.

\bibitem{newlook} M. Talagrand 
{\sl A new look at independence}, Ann. Probab. {\bf 24}, 1 (1996).

\bibitem{talaconc} M. Talagrand, 
{\sl Concentration of measure and 
isoperimetric inequalities in product spaces}, 
Publ. Math. I.H.E.S. {\bf 81}, 73 (1995).

\bibitem{shiri}  A. N. Shiryaev, {\sl Probability}, Springer, Berlin (1989).

\bibitem{noinuovo} F. Guerra, F. L. Toninelli, in preparation

\end{thebibliography}
\end{document}